\documentstyle[sprocl,psfig]{article}

\def\Journal#1#2#3#4{{#1} {\bf #2}, #3 (#4)}

\def\NPB{{\em Nucl. Phys.} B}
\def\PLB{{\em Phys. Lett.}  B}
\def\PRL{\em Phys. Rev. Lett.}
\def\PRD{{\em Phys. Rev.} D}

\def\be{\begin{equation}}
\def\ee{\end{equation}}
\def\bea{\begin{eqnarray}}
\def\eea{\end{eqnarray}}

\newcommand{\bref}[1]{(\ref{#1})} 
\newcommand{\sGUT}{{\scriptscriptstyle \rm GUT}}

 	\def\GeV{\,\mbox{GeV}}
 		
 \def\Ev{E_{\rm v}}		  
 		
 \def\rhv{\rho_{\rm v}}		    
 		
 \def\Tx{T_{\rm x}}		\def\Ts{T_{\rm s}}
 			
 \def\TGUT{T_\sGUT}		
\newcommand{\mt}{\widetilde m}
\newcommand{\mB}{m_{\scriptscriptstyle \rm B}}
\newcommand{\ms}{m_{\rm s}}
\newcommand{\mPl}{m_{\scriptscriptstyle \rm Pl}}

\begin{document}

\title{The Fate of Cosmic String Zero Modes}

\author{A. C. DAVIS}
\address{Department of Applied Mathematics and Theoretical Physics,\\
	 Centre for Mathematical Sciences,\\
	 University of Cambridge, Cambridge, CB3 0WA, UK}
\author{S. C. DAVIS and W. B. PERKINS}
\address{Department of Physics, University of Wales Swansea,\\
Singleton Park, Swansea, SA2 8PP, UK}


\maketitle\abstracts{ 
              Cosmic strings in realistic particle physics models 
              have fermion zero modes in the string core. Such
              zero modes change the underlying cosmology; for
              example, the string can carry a current. We consider
              the stability of fermion zero modes to interactions with
              particles in the surrounding plasma. We also consider
              cosmological constraints on such theories, particulary
              those with a chiral fermion zero mode, such as arises    
              in a class of supersymmetric models.}

\section{Introduction}

Although much of the evolution of the Universe is well understood,
there are still many cosmological phenomena for which a completely
satisfactory explanation has yet to be found. Topological defects, such
as cosmic strings, could provide mechanisms for structure formation,
CMB anisotropy, and high energy cosmic rays~\cite{rays}. Such defects form
in many realistic particle physics theories, including those involving
supersymmetry. 

In the past it has been difficult to evaluate the usefulness of such
ideas due to a lack of data. This is now changing, and predictions of
CMB anisotropies from simple cosmic string models have been
made~\cite{Turok}. While these predictions show poor agreement with
the observations, they  do not take into account the full physics of
string models. Indeed, recent analysis which includes the effect of
particle production as an energy loss mechanism from the string network
shows much improved agreement with data~\cite{joao&nathalie}.

One significant possibility is that the strings carry conserved
currents~\cite{supercond}. These currents will alter the evolution of
a string network, which could lead to better agreement with
observation. Indeed, an analytic analysis showed that a much denser 
string network results for electromagnetically coupled strings~\cite{kostas}.
However, one significant implication of conserved currents
is that they can stabilise loops of string. If persistent,
these stabilised loops or `vortons'~\cite{vorton,vorton2}, can easily
dominate the energy density of the Universe, placing stringent
constraints on the parameters of the model. An analysis has been made of
the implications of this for particle physics models that predict current
carrying strings~\cite{brandon&acd1,brandon&acd2}.

Fermions are a natural choice for the charge carriers of such
currents. Fermion zero modes exist in a wide class of cosmic string
models. The fermions can be  excited and move along the string,
resulting in a current. Consequently,  fermion conductivity occurs
naturally in many supersymmetric~\cite{DDT} and grand unified
theories~\cite{so10EWstr}, such as SO(10). In particular, in SUSY
models with a D term, there is a single chiral zero mode, either a
left or right mover. This zero mode survives  supersymmetry
breaking~\cite{DDT2}. This class of theory is potentially important since it
arises in many superstring models.

One criticism of fermion currents is that, unlike scalar boson
currents, they are not topologically stable. It is possible that they
could decay, either directly into particles off the string, or through
interactions with the surrounding plasma. If the decay rate is too
high, currents will not last long enough to have any significant
effect. On the other hand if the decay rate is too low the Universe
could become vorton dominated. We address this issue and show that 
these processes can remove current carriers
close to the phase transition, but not otherwise~\cite{DPD}. 
We then consider vorton constraints in chiral theories, showing this 
leads to stringent constraints on the underlying particle physics theory.

\section{$D$-term Supersymmetric Cosmic Strings}
\label{SUSY sec}

Consider a supersymmetric theory with a $\mbox{U(1)} \rightarrow I$
phase transition. The simplest way to achieve this symmetry breaking
is to use one charged chiral superfield and a non-zero Fayet-Iliopoulos
term. Expanding the Lagrangian in terms of component fields gives (in
Wess-Zumino gauge)
\bea
&&{\cal L} = |(D_\mu \phi|^2 - \frac{1}{4} F^{\mu\nu}F_{\mu\nu} + |F|^2
	+ \frac{1}{2}D^2 + D (g|\phi|^2 + \xi)
\nonumber\\ &&\hspace{.5in}- i\psi \sigma^\mu D^{\ast}_\mu \bar \psi 
- i\lambda \sigma^\mu \partial_\mu \bar\lambda
+ ig\sqrt{2}\phi^* \psi \lambda + (\mbox{c.c.}) \ .
\eea

The two fermion fields $\psi$ and $\lambda$ are the Higgsino and
gaugino respectively.  Eliminating the two auxillary fields $D$ and
$F$ will give potential terms. If $\xi = - g \eta^2 < 0$ the resulting
$D$-term will give rise to spontaneous symmetry breaking, resulting
in an expectation value of $\eta$ for $\phi$. In order to avoid gauge
anomalies, the model must contain other charged superfields. We will
assume these have zero expectation values.

As well as the $\phi=$ constant solution there also
exist string solutions obtained from the ansatz
\begin{eqnarray} 
\phi & = & \eta e^{in\theta}f(r) \\
A_\mu & = & - n \frac{a(r)}{gr}\delta_\mu^\theta \\
D & = & g\eta^2 (1-f(r)^2) \ ,
\end{eqnarray}
where$f(r)$ and $a(r)$ are the usual Neilsen-Olesen 
profile functions.

Now consider the fermionic sector of the theory. Performing a SUSY
transformation with 2-spinor parameter $\epsilon$ gives
\bea 
\delta \lambda_1 &=& 2 i g \eta^2(1-f^2) \epsilon_1 \nonumber \\
\delta \psi_1 &=& -2\sqrt{2} i\eta \frac{n}{r}(1-a)f 
			e^{i(n-1)\theta}\epsilon_1^* \ .
\label{susyzm} 
\eea
$\delta\psi_2$ and $\delta\lambda_2$ are both zero. The
expression \bref{susyzm} is a zero energy solution of the fermion
field equations. It is trivial to add $z$ and $t$ dependence to it, and
the resulting solution is a null current moving along the string.

Since the string solution is not invariant under transformations with
$\epsilon_1 \neq 0$, supersymmetry has been broken inside the
string core. However it is only partially broken there since the string is
still invaraint under transformations parametrised by $\epsilon_2$.  

It is also possible to get a phase transition using an $F$-term. This
requires a non-trivial superpotential and at least three scalar
fields. The corresponding string solutions have twice as many zero
modes as the $D$-term theory, and move in both directions along the
string. In this case supersymmetry is completely broken~\cite{DDT}. In both 
theories the zero modes and SUSY breaking are confined to the string core.

When soft supersymmetry-breaking terms are included in the Lagrangian,
oppositely moving zero modes in the $F$-term theory mix to form
massive states~\cite{DDT2}. This cannot happen in the $D$-term theory since the
zero modes all move in the same direction.

Thus, a generic property of cosmic strings in SUSY theories is that 
supersymmetry is broken in the string core and the resulting strings have
fermion zero modes. As a consequence, cosmic strings arising in SUSY theories
are automatically current-carrying and can give rise to vortons.

\section{Decay Rates of Charge Carriers}

Vortons can only be stable if the currents they carry are stable, thus
the stabilty of the charge carriers is a crucial consideration. In
the absence of other particles, currents carried by zero modes on
isolated, straight strings are stable on grounds of energy and
momentum conservation. However, in a realistic setting there are many
processes which can depopulate zero modes. Strings are not isolated
and in the early Universe they and their bound states will interact
with the hot plasma. SO(10) is considered as a specific example. In
this case a heavy neutrino zero mode may scatter from a light plasma
particle to produce a light fermion-antifermion pair via an
intermediate electroweak Higgs boson. A massive neutrino bound mode may also
simply decay into light fermions. Further, bound states on different
strings, or different parts of a single curved string, can scatter
from one another by exchanging a Higgs particle.

The decay rates of both massive and massless modes have been
calculated at tree level~\cite{DPD}. The main difference between
scattering in the string background and that in a trivial background
is the lack of  translation invariance transverse to the string. This
enters the calculation via the bound mode wavefunctions which are
localised around the string. Integration over the initial vertex
position does not yield the standard momentum conserving
$\delta$-function, instead it produces an approximately gaussian
function that permits non-conservation of transverse momentum on the
scale of the off-string fermion mass.

The violation of momentum conservation in the string background opens
up significant areas of phase space that are forbidden in the trivial
background. Of particular importance is the possibility of resonant
scattering. For example the lifetime  of a bound mode of mass $\mB$ 
decaying into a light fermion and an electroweak Higgs is,
\be
\tau \sim (|g_\nu|^2 \mB)^{-1} \ ,
\label{mb life}
\ee
where $g_\nu$ is the Yukawa coupling in the neutrino's electroweak
mass term. Massive modes with the appropriate  couplings to the
electroweak sector thus have a very short lifetime. The above result
neglects the possibility that incoming and outgoing fermionic
wavefunctions may be amplified near the string~\cite{wfnamp} and
should be taken as an overestimate. 

The fate of massless bound states is also complicated by momentum
non-conservation. Of particular interest are the bound states that
stabilise vortons. Contraction of the vorton will ensure that the
states at the Fermi surface will have GUT scale momenta. The lifetime
of these high momentum states is critical; if they decay the vorton
will contract and {\it promote} low momentum states to high
momentum. Energy and $z$-momentum conservation prevent massless states
from decaying spontaneously. However it is possible for them to decay
by interaction with plasma particles or other zero modes. If the
string Higgs mass, $\ms$  exceeds the off-string fermion mass, $m$, it
is also possible for high energy massless currents on a curved string
to decay by tunneling to free heavy neutrinos~\cite{vorton}. However
the rate will not be significant unless $\ms \gg m$.

If we consider a massless bound state at the Fermi surface scattering
from a typical plasma particle, we have a centre of mass energy of
order $\sqrt{\ms T}$. In the SO(10) case this is well above the mass
of the electroweak Higgs intermediate particle and  transverse
momentum non-conservation again allows for resonant scattering. 
Including amplification of the incoming plasma particle wavefunction,
the lifetime of these high momentum zero modes is found to be,
\be
\tau \sim {\mt^2\over |g_\nu|^2 T^3}\left({T\over \mt}\right)^{2Q} \ ,
\ee
where $\mt=\ms m/2$ and  $Q$ is the charge of the plasma particle
under the string gauge field. 

In the radiation dominated era the time is given by $t=\alpha T^{-2}$,
where $\alpha \sim \mPl/10$. In the case of SO(10), $Q=3/10$,
and the probability of a zero mode state scattering
after some time $t_i$ is small if,
\be
t_i > O\left([\mt^{-7} |g_\nu|^{10} \alpha^{6}]\right) > 
O\left(\left[{\mPl\over 10\mt}\right]^6 |g_\nu|^{10}\mt^{-1}\right) \ .
\ee  

As the lifetime varies only slightly faster than $T^{-2}$, this result
for $t_i$ is very sensitive to the Yukawa coupling. For $\mt \sim
10^{15}$GeV, if $|g_\nu|=1$, zero mode states populated after $t_i\sim
10^{15}t_\sGUT$ will be stable, while if $|g_\nu| < 0.03$, this
scattering is never significant. In the SO(10) model $g_\nu$ is also
the Yukawa coupling for the corresponding quarks, thus there is an
epoch when $\nu^c_\tau$ zero modes will scatter from the string, but
$\nu^c_e$ and $\nu^c_\mu$ zero modes will never scatter by this
process. Thus the interaction with plasma particles can not significantly
remove zero modes from the string.  
Note that it is also possible to create currents using the
above interactions in reverse. Hence, if thermal equilibrium is reached the
number density of zero modes will be of order $T$.

Within the SO(10) model there is also the possibility of mediating
these processes by GUT mass Higgs fields with zero VEV. In this case
the Yukawa coupling need not be small, but the centre of mass energy
of the interaction is only of order the intermediate particle mass for
$T\sim \TGUT$. Thus below the GUT temperature the reaction rates for
these processes are rapidly suppressed by powers of $T/\TGUT$.

None of these plasma scattering processes can remove $\nu^c_e$ and
$\nu^c_\mu$ zero modes, and are only significant for $\nu^c_\tau$ immediately
after the phase transition. Thus, they are unable to prevent the vorton
density from dominating the energy density of the Universe.

The plasma scattering processes considered above failed to remove zero
modes due to the decreasing plasma density at late times. A distinct
category of process is the scattering of a zero mode on one string by
a second zero mode on another string. This is particularly relevant
for vortons as they form small loops with a typical radius only one or two
orders of magnitude larger than the string width. We thus have zero
modes moving in opposite directions on opposite sides of the
vorton. For simplicity the decay rate can be calculated by considering
two straight, anti-parallel strings~\cite{DPD} with spacing $2R$. 

If the intermediate particle is an electroweak Higgs boson, 
$R$ is much smaller than the electroweak length scale and there is
no exponential {\it range} suppression. Resonant scattering is not
possible and the cross-section is found to be,
\be
\sigma\sim {|g_\nu|^4\over (m_\sGUT R)^4} \ .
\label{vor sigx}
\ee
This cross-section is dimensionless as the scattering is effectively
in one spatial  dimension. 
 
Conversely, for a GUT mass intermediate particle, 
$m_\sGUT R\sim 10-100$ and  the reaction rate displays
exponential {\it range} suppression,
\be
\sigma\sim {|g_\nu|^4 \over (m_\sGUT R)^3} e^{-4 m_\sGUT R} \ .
\label{vor sigG}
\ee
While the string density may be high at formation, it drops rapidly
and the exponential suppression in \bref{vor sigG} makes such processes
irrelevant in all  physical situations.

Taken at face value, if electroweak particles mediate current--current
scattering on different segments of string then \bref{vor sigx} gives
a short lifetime for charge carriers on a vorton. However, for a circular 
loop the angular momentum of 
the fermions  must be conserved and, combined with energy
conservation, this would prevent massless modes on the string
scattering into massive modes. However, the fermionic spectrum has not
been calculated for a circular loop, so there is no reason to expect
the zero modes to remain massless. Thus angular momentum
conservation can only be considered in a consistent, rotationally
invariant calculation. The calculation above works consistently with
straight strings, thus while \bref{vor sigx} may not be directly
applicable to vorton decay, it is relevant for interactions on
non-circular loops and scattering of currents on a string network.

\section{Chiral Strings and Vortons}

We have seen that plasma interactions donot significantly remove zero
modes from the string. This is particularly true in SUSY $D$-term 
theories (see section~\ref{SUSY sec})
where we would expect the zero mode to be isolated from the electroweak
sector. As a consequence, stable vortons would be expected
to form. In this section we constrain the underlying particle physics
theory which gives rise to chiral zero modes on the string.
We assume that the strings are formed at a phase transition
ocurring at temperature $\Tx$ and become current-carrying at a scale $\Ts$.
The string loop is characterised by two currents, the topologically conserved
phase current $N$ and the dynamically conserved particle number current
$Z$. In the chiral case these are exactly identical.

A non conducting string loop must ultimately decay by radiative and
frictional drag processes until it disappears completely. However, a
conducting string loop could reach a state in which the energy attains
a minimum for given non zero values of $N$ and $Z$. This is the vorton
state.  It should be emphasised that the existence of such vorton
states does not require that the carrier field be electromagnetically
coupled. Indeed, in the case of the $D$-term zero modes, it is not.

The physical properties of a vorton state are determined by the quantum
numbers, $N$ and $Z$. However, these are not arbitrary. Indeed, to
avoid the fate of the usual loops, the quantum numbers on a conducting loop 
must be large compared with unity. This in turn implies
a minimum length for the loop. In our work we calculate the number density
of protovorton loops, subject to the loops being greater than a minimum
length. We then estimate the vorton density and constrain the underlying
theory by the requirement that the universe is not vorton dominated. We
apply two constraints. Firstly we take a conservative assumption that the
vortons only live a few minutes and constrain the theory by requiring that
the universe is radiation dominated at nucleosynthesis; we then take the
more realistic assumptions that chiral vorton loops are stable and 
constrain the theory by requiring that the vorton density is less than the
closure density today. The full details are presented in 
ref.~\cite{brandon&acd2}; here we sumarise the main results.

For chiral vortons the vorton energy, $\Ev$, is given by
\be
\Ev\simeq N \Tx \ . 
\ee

In the friction dominated era the number density of vortons can be estimated
by considering the damping length scale and the resulting correlation length
below which microstructure is damped. This automatically satisfies the
minimum length criteria mentioned above. For a loop of length L,
the conserved quantum number is then
\be
\vert Z\vert = N \approx {L \Ts} = 
		\left({\mPl \over \Ts}\right)^{1/2} {\Tx\over\Ts} \ .
\ee

The resulting vorton density was found to be
\be
{\rhv\over T^3} \approx {\Ts^{3} \over \mPl \Tx} \ ,
\ee
where factors of order unity have been dropped.

Requiring then that this vorton density be less than the radiation
density at nucleosynthesis gives a constraint for strings which become 
current-carrying at formation of
\be
\Tx\leq 10^8 \GeV \ .
\ee

This is the condition that must be satisfied by the formation
temperature of {\it cosmic strings that become current-carrying
immediately}, subject to the rather conservative assumption that the
resulting vortons last for at least a few minutes. 

If the zero modes condense on the string at a separate phase
transition then the constraint takes us outside the friction dominated
regime. The calculation is much more involved and we refer the reader
to ref.~\cite{brandon&acd2} for details. However, the resulting
constraint is that GUT scale strings becoming current-carrying at a
temperature above $10^9 \GeV$ are inconsistent with data. 

Being less conservative we can make the assumption that the vortons are
absolutely stable and survive to the present time. Requiring that the
vorton density is less than the closure density gives a stronger constraint.
In this case we find
\be
\Tx = \Ts \leq 10^5 \GeV \ .
\ee

For strings which become current-carrying at a later transition, the 
details are again more complicated. However, for GUT scale strings 
we find that strings becoming current-carrying at a temperature above
$10^5 \GeV$ are inconsistent with data.

It is amusing to point out that if strings formed, or became
current-carrying, just below this temperature, then their vortons
would contribute substantially to the dark matter of the universe.

\section{Discussion}

We have shown that cosmic strings arising in SUSY theories are generically
current-carrying. For D-term theories there is a chiral fermion zero mode
which survives SUSY breaking. Consequently, the current persists.

We have investigated ways in which the fermion bound modes could be
destabilised. Zero modes are the most resilient: various scattering
mechanisms were investigated, but, because of the low density of the
surrounding plasma, they were unable to scatter all zero modes off the
string. Zero modes on neighbouring string segments can also scatter
off each other. This may be relevant for vortons. 

Strings with a chiral zero mode lead to the most stringent constraints on
the underlying particle physics theory, because in this case the fermion
travels in a single direction, resulting in the current being maximal
rather than random. We have shown that this leads to very strong constraints
indeed.

However, if the theory somehow manages to evade the vorton constraints then
the resulting cosmology could be very different. For example, the scaling
solution of this type of cosmic string theory is completely unknown. Whilst
the cosmology of strings resulting from the abelian Higgs model has been
investigated, that resulting from realistic GUT theories has not. 
As a consequence. it is premature to rule out cosmic strings as being 
incompatible with microwave background data. This is currently under
investigation~\cite{Dani}.

\section*{Acknowledgments}
This work was supported in part by PPARC and an ESF network. We wish to
thank our collaborators Brandon Carter and Mark Trodden, and also Rachel
Jeannerot and the other organisers of COSMO99.

\end{document}